\documentclass[amsmath,amssymb]{revtex4}

\usepackage{graphicx}
\usepackage{dcolumn}
\usepackage{bm}

\def\q1{{q^{-1}}}
\def\dj{{{\cal D}_x}}

\def\qf2{{q^{1/2}}}
\def\qmf2{{q^{-1/2}}}

\newcommand{\be}{\begin{equation}}
\newcommand{\ee}{\end{equation}}
\newcommand{\beq}{\begin{eqnarray}}
\newcommand{\eeq}{\end{eqnarray}}

\begin{document}
\title{Basic-deformed quantum mechanics}
\author{A. Lavagno}
\affiliation{Dipartimento di Fisica, Politecnico di Torino,
I-10129 Torino, Italy
\\ Istituto Nazionale di Fisica Nucleare (INFN),
Sezione di Torino, Italy }

\begin{abstract}
Starting on the basis of $q$-symmetric oscillator algebra and on
the associate $q$-calculus properties, we study a deformed quantum
mechanics defined in the framework of the basic square-integrable
wave functions space. In this context, we introduce a deformed
Schr\"odinger equation, which satisfies the main quantum mechanics
assumptions and admits, in the free case, plane wave functions
that can be expressed in terms of the $q$-deformed exponential,
originally introduced in the framework of the basic-hypergeometric
functions.
\end{abstract}
\maketitle

\section{Introduction}
Quantum algebras are deformed version of the Lie algebras, to
which they reduce when the deformation parameter $q$ is set equal
to unity. Their use in physics became popular from the works of
Biedenharn \cite{bie} and Macfarlane \cite{mac} and with the
introduction of the $q$-deformed harmonic oscillator based on the
construction of $SU_q(2)$ algebra of $q$-deformed commutation or
anticommutation relations between creation and annihilation
operators. Initially used for solving the quantum Yang-Baxter
equation, quantum algebra have subsequently found several
applications in different physical fields such as cosmic strings
and black holes \cite{strominger}, conformal quantum mechanics
\cite{youm}, nuclear and high energy physics \cite{bona,svira},
fractional quantum Hall effect and high-$T_c$ superconductors
\cite{wil}. At same time, it was clear that the $q$-calculus,
originally introduced in the study of the basic hypergeometric
series \cite{jack,exton,gasper}, plays a central role in the
representation of the quantum groups with a deep physical meaning
and not merely a mathematical exercise.

In Ref.s \cite{pre,qkinetic}, it was shown that a natural
realization of quantum thermostatistics of $q$-deformed bosons and
fermions can be built on the formalism of $q$-calculus. In Ref.s
\cite{torun,epjc06}, a $q$-deformed Poisson bracket, invariant
under the action of the $q$-symplectic group, has been derived and
a classical $q$-deformed thermostatistics has been proposed in
Ref.\cite{noi2}. It is further remarkable to point out that
relativistic mechanics in the two-dimensional noncommutative
Minkowski space-time has been introduced in Ref. \cite{rembie},
$q$-deformed Hamilton equations of motion have been studied in
Ref. \cite{caban}, while, in Ref. \cite{chung}, $q$-deformed
classical mechanics has been derived by using the variational
technique on a $q$-deformed Lagrangian.

Furthermore, it has been shown that it is possible to obtain a
"coordinate" realization of the Fock space of the $q$-oscillators
by using the Jackson derivative (JD) \cite{flo,fink,cele}. In
Ref.\cite{cele}, the quantum Weyl-Heisenberg algebra is studied in
the frame of the Fock-Bargmann representation and is incorporated
into the theory of entire analytic functions allowing a deeper
mathematical understanding of squeezed states, relation between
coherent states, lattice quantum mechanics and Block functions.
Moreover, in the recent past, a tentative to constructing a
classical counterpart to the quantum group and $q$-deformed
dynamics have been investigated \cite{marmo1}. In this context, it
appears important to outline that in Ref.s
\cite{kowa1,kowa2,kowa3} a very general approach, based on
Bargmann representation in connection to coherent states, has been
introduced.

It is remarkable to observe that $q$-calculus is very well suited
for to describe fractal and multifractal systems. As soon as the
system exhibits a discrete-scale invariance, the natural tool is
provided by JD and $q$-integral in the framework of the basic
hypergeome\-tric functions, which constitute the natural
generalization of the regular derivative and integral for
discretely self-similar systems \cite{erzan1}. It is also relevant
to outline that the statistical origin of such $q$-deformation
lies in the modification, relative to the standard case, of number
of states $W$ of the system corresponding to the set of
occupational number ${n_i}$ \cite{pre}. In literature, other
statistical generalization are present, such as the so-called
nonextensive thermostatistics or superstatistics with a completely
different origin \cite{tsallis,pla02,plb01}.

In the past, the study of generalized linear and non-linear
Schr\"odinger equations has attracted a lot of interest because of
many collective effects in quantum many-body models can be
described by means of effective theories with generalized
one-particle Schr\"odinger equation
\cite{pita,gross,doebner,suther}. In Ref. \cite{wess2} a
deformation of the canonical commutation relations, based on
quantum group arguments, has been studied and in Ref.
\cite{kozyrev} the free motion of $q$-deformed quantum particle
has been investigated. Always in the framework of the
$q$-Heisenberg algebra, $q$-deformed Schr\"odinger equations have
been proposed \cite{Zhang,Micu,dobro}.

Although the inve\-sti\-gated quantum dynamics is based on
noncommutative differential structure on configuration space, we
believe that a fully consistent $q$-deformed formalism of the
quantum dynamics, starting from the properties of $q$-calculus and
basic-hypergeometric functions, has been still lacking.
Furthermore, it is known in literature that in the framework $q$
non-symmetric deformed algebra, the momentum operator, associate
to the JD, is not Hermitian in the common sense and a complicate
combination of JD must be postulated to introduce a $q$-deformed
quantum mechanics \cite{wess,annaby,lava08}.

In this paper, working in the context of the $q\leftrightarrow\q1$
symmetric algebra, we study a generalized deformed quantum
mechanics consistently with the prescriptions of the $q$-calculus.
At this scope, we introduce a scalar product in a basic
square-integrable wave space and an associate definition of
basic-adjoint and basic-Hermitian operators. In this framework, we
introduce a $q$-deformed Schr\"odinger equation which satisfies
the principal assumptions of the quantum mechanics.

As mentioned before, we like to outline that in literature exist
several examples of concrete functions representation of deformed
quantum mechanics, that are different from our approach, based on
$q$-calculus and basic hypergeometric functions. Besides the
already quoted papers, very recently coherent states for
$q$-deformed quantum mechanics on a circle have been rigorously
investigated \cite{kowa4}.

The paper is organized as follows. In Sec. 2, we review the main
features, useful in the present investigation, of $q$-deformed
oscillator algebra and the principal properties of $q$-calculus
and basic hypergeometric elementary functions. In Sec. 3, we
introduce the principal operator properties in the so-called basic
square-integrable space, well known in mathematical literature of
the basic hypergeometric functions. These properties allow us to
introduce, in Sec. 4, the mean value of a dynamical variable and a
generalized basic-deformed Schr\"odinger equation which admits, in
the free case, plane wave functions that can be expressed in terms
of the basic-hypergeometric exponential function. A brief
conclusion is reported in Sec. 5.

\section{Deformed oscillator algebra, $q$-calculus and basic elementary functions}
We shall review the principal relations of $q$-oscillators defined
by the $q$-Heisenberg algebra of creation and annihilation
operators introduced by Biedenharn and McFarlane \cite{bie,mac},
derivable through a map from SU$_q$(2). Furthermore, we will
review the main features of the strictly connected $q$-calculus
and basic-deformed elementary functions useful in the present
investigation. Although most of the collected material and results
reported in this Section are not original, this review constitutes
the basis of the following study and represents a bridge between
physical and mathematical concepts known in literature.

The symmetric deformed algebra is determined by the following
commutation relations for $a$, $a^{\dag}$ and the number operator
$N$, thus (for simplicity we omit the particle index)
\begin{equation}
[a,a]=[a^\dag,a^\dag]=0 \; , \ \ \ aa^\dag-q a^\dag a =q^{-N} \; ,
\label{algebra}
\end{equation}
\begin{equation}
[N,a^\dag]= a^\dag \; , \ \ \ [N,a]=-a\; .
\end{equation}

The $q$-Fock space spanned by the orthornormalized eigenstates
$\vert n\rangle$ is constructed according to
\begin{equation}
\vert n\rangle=\frac{(a^\dag)^n}{\sqrt{[n]!}} \vert 0\rangle \; , \
\ \ a\vert 0\rangle=0 \; , \label{fock}
\end{equation}
where the $q$-basic factorial is defined as
\begin{equation}
[n]!=[n] [n-1] \cdots [1] \, ,\label{brnf}
\end{equation}
and the so-called $q\leftrightarrow\q1$ symmetric basic number $[x]$
is defined in terms of the $q$-deformation parameter
\begin{equation}
[x]=\frac{q^{x}-q^{-x}}{q-\q1}\; . \label{bn}
\end{equation}

In the limit $q \rightarrow 1$, the basic number $[x]$ reduces to
the ordinary number $x$ and all the above relations reduce to the
standard boson relations.

The actions of $a$, $a^\dag$ on the Fock state $\vert n \rangle$ are
given by
\begin{eqnarray}
a^\dag \vert n\rangle &=& [n+1]^{1/2} \vert n+1\rangle\; , \\
a \vert n\rangle&=&[n]^{1/2} \vert n-1\rangle \; ,\\
N \vert n\rangle&=&n\vert n\rangle \; .
\end{eqnarray}
From the above  relations, it follows that $a^\dag a=[N]$,
$aa^\dag=[N+1]$.

The transformation from Fock observable to the configuration
space (Bargmann holomorphic representation) may be accomplished by
the replacement \cite{flo,fink}

\begin{equation}
a^\dag\rightarrow x \; , \ \ \ a\rightarrow \dj \; , \label{jd}
\end{equation}
where $\dj$ is the Jackson derivative (JD) \cite{jack}
\begin{equation}
\dj=\frac{D_x-(D_x)^{-1}}{(q-\q1)\,x} \ ,
\end{equation}
and
\begin{equation}
D_x=q^{x\,\partial_x} \, , \label{dila}
\end{equation}
is the dilatation operator. Its action on an arbitrary real
function $f(x)$ is given by
\begin{equation}
\dj\,f(x)=\frac{f(q\,x)-f(\q1 x)}{(q-\q1)\,x} \ .
\end{equation}
In contrast to the usual derivative, which measures the rate of
change of the function in terms of an incremental translation of its
argument, the JD measures its rate of change with respect to a
dilatation of its argument by a factor of $q$.

The JD satisfies some simple proprieties which will be useful in the
following. For instance, its action on a monomial $f(x)=x^n$, where
$n\geq0$, is given by
\begin{equation}
\dj\,(a\,x^n)=a\, [n]\,x^{n-1} \ ,\label{7}
\end{equation}
where $a$ is a real constant.

Moreover, it is easy to verify the following basic-version of the
Leibnitz rule
\begin{eqnarray}
\nonumber \dj\Big(f(x)\,g(x)\Big)&=&\dj\,f(x)\,g(\q1 x)+f(q\,x)\,\dj\,g(x) \ ,\\
&=&\dj\,f(x)\,g(q\,x)+f(\q1 x)\,\dj\,g(x) \ .\label{leib}
\end{eqnarray}
In addition the following property holds
\begin{eqnarray}
{\cal D}_{ax}\,f(x)=\frac{1}{a}\,\dj f(x) \ .\label{chain}
\end{eqnarray}

Consistently with the $q$-deformed theory, the standard integral
must be generalized to the basic-integral defined, for $0<q<1$ in
the interval $[0,a]$, as \cite{bona,exton}
\begin{equation}
\int_0^a f(x)\, d_q x=a\,(\q1-q)\, \sum_{n=0}^\infty q^{2n+1}\,
f(q^{2n+1}\,a) \, ,
\end{equation}
while in the interval $[0,\infty)$
\begin{equation}
\int_0^\infty f(x)\, d_q x=(\q1-q)\, \sum_{n=-\infty}^\infty
q^{2n+1}\, f(q^{2n+1}) \, .
\end{equation}
The indefinite $q$-integral is defined as
\begin{equation}
\int f(x)\, d_q x=(\q1-q)\, \sum_{n=0}^\infty q^{2n+1}\,x\,
f(q^{2n+1}\,x)+{\rm constant} \, .
\end{equation}
One can easily see that the basic-integral approaches the Riemann
integral as $q\rightarrow 1$ and also that $q$-differentiation and
$q$-integration are inverse to each other, thus
\begin{equation}
\dj\int_0^x f(t)\, d_q t=f(x) \, , \ \ \ \int_0^a \dj f(x)\, d_q
x=f(a)-f(0) \, ,
\end{equation}
where the second identity occurs when the function $f(x)$ is
$q$-regular at zero, i.e.
\begin{equation}
\lim_{n\rightarrow \infty} f(xq^n)=f(0) \, .
\end{equation}

By using the deformed Leibnitz rule of Eq.(\ref{leib}), analogue
formulas for integration by parts may easily be deduced as
\begin{eqnarray}
\int_0^a f(qx)\, \dj g(x)\, d_q
x=f(x)\,g(x)\vert^{x=a}_{x=0}-\int_0^a \dj f(x)\, g(\q1 x) \, d_q
x \,  , \\
\int_0^a f(\q1x)\, \dj g(x)\, d_q
x=f(x)\,g(x)\vert^{x=a}_{x=0}-\int_0^a \dj f(x)\, g(q x) \, d_q x
\,  .
\end{eqnarray}
For the following developments it appears very relevant to observe
that the above integration by parts can be also expressed, for
example, as
\begin{eqnarray}
\int_0^a f(x)\, \dj g(x)\, d_q x=f(q
x)\,g(x)\vert^{x=a}_{x=0}-\int_0^a  \dj f(qx)\, g(q x) \, d_q x \,
, \label{leibq}
\end{eqnarray}
or, equivalently,
\begin{eqnarray}
\int_0^a f(x)\, \dj g(x)\, d_q x=f(\q1
x)\,g(x)\vert^{x=a}_{x=0}-\int_0^a  \dj f(\q1 x)\, g(\q1 x) \, d_q
x \, ,\label{leibq1}
\end{eqnarray}
where we have rewritten the $q$-Leibniz rule  as
\begin{eqnarray}
\dj\Big(f(qx)\,g(x)\Big)=\dj\,f(qx)\,g(qx)+f(x)\,\dj\,g(x) \, ,
\end{eqnarray}
or
\begin{eqnarray}
\dj\Big(f(\q1x)\,g(x)\Big)=\dj\,f(\q1 x)\,g(\q1 x)+f(x)\,\dj\,g(x)
\, .
\end{eqnarray}

From the above relations, as also pointed out in
Ref.\cite{erzan1,sornette}, it appears evident that JD and
$q$-calculus provides a custom made formalism in which to express
scaling relations. When $x$ is taken as the distance from a
critical point, JD thus quantifies the discrete self-similarity of
the function $f(x)$ in the vicinity of the critical point and can
be identified with the generator of fractal and multifractal sets
with discrete dilatation symmetries.

Let us now introduce the following $q$-deformed exponential function
defined by the series
\begin{equation}
{\rm E}_q(x)= \sum_{k=0}^{\infty}\, \frac{x^k}{[k]_q!}=1+x+
\frac{x^2}{[2]_q!}+\frac{x^3}{[3]_q!}+\cdots \ .\label{qexp}
\end{equation}
The function (\ref{qexp}), reducing to the ordinary exponential
function in the $q\to1$ limit, defines the basic-exponential, well
known in the literature since a long time ago, originally introduced
in the study of basic hypergeometric series \cite{exton,gasper}.

In this context, it is relevant to observe that the
basic-exponential can be also written as
\begin{equation}
{\rm E}_q (x)= \sum_{n=0}^{\infty}\,
\frac{(1-q)^n\,x^n}{(q;q)_n}\,q^{n(n-1)/2}\ ,\label{qexpmath}
\end{equation}
where we have introduced the $q$-shifted factorial, defined by
\begin{equation}
(a;q)_0=1 \, , \ \ \ \ (a;q)_n=\prod_{k=0}^{n-1} (1-aq^k)\, , \ \ \
(a;q)_\infty=\prod_{k=0}^\infty (1-aq^k)\, ,
\end{equation}
and we have used the identity
\begin{equation}
\frac{(1-q)^n}{(q;q)_n}\,q^{n(n-1)/2}=\frac{1}{[n]_q!} \, .
\label{qfattpr}
\end{equation}
The right hand expression of Eq.(\ref{qexpmath}) occurs
principally in mathematical literature, being the $q$-shifted
factorial commonly used in the framework of the basic
hypergeometric functions \cite{gasper}.

Among many properties, it is important to recall the following
relation \cite{exton}
\begin{equation}
\dj {\rm E}_q(a\,x)= a\,{\rm E}_q(a\,x) \ ,\label{jde}
\end{equation}
and its dual
\begin{equation}
\int\limits_0\limits^x{\rm E}_q(a\,y)\,d_qy=\frac{1}{a}\,\Big[{\rm
E}_q(a\,x)-1\Big] \, ,\label{dae}
\end{equation}
where $a$ is a real number different from zero. It is important to
point out that Eqs.(\ref{jde}) and (\ref{dae}) are two important
properties of the {\em basic}-exponential which turns out to be
not true if we employ the ordinary derivative or integral.

Beside to the $q$-deformed exponential, it is natural to introduce
the basic-deformed trigonometric functions as \cite{exton}
\begin{equation}
{\rm E}_q(ix)=C_q(x)+i\,S_q(x)  \, . \label{trigq}
\end{equation}
As a consequence, we have
\begin{eqnarray}
&&S_q(x)= \sum_{n=0}^{\infty}\, (-1)^n\,
\frac{x^{2n+1}}{[2n+1]_q!} \, , \label{qsin}\\
&&C_q(x)= \sum_{n=0}^{\infty}\, (-1)^n\, \frac{x^{2n}}{[2n]_q!} \,
,\label{qcos}
\end{eqnarray}
and it easy to see that
\begin{eqnarray}
&&S_q(x)=[{\rm E}_q(ix)-{\rm E}_q(-ix)]/2i \, ,\\
&&C_q(x)=[{\rm E}_q(ix)+{\rm E}_q(-ix)]/2 \, .
\end{eqnarray}
Analogously to Eq.(\ref{qexpmath}), the basic-trigonometric
functions can be cast in term of the $q$-shifted factorial as
\begin{eqnarray}
&&S_q(x)=\sum_{n=0}^{\infty}\,(-1)^n\,\frac{q^{n(n+1/2)}}{(1-q)\,(q^2,q^3;q^2)_n}\,
x^{2n+1} \, ,\label{qsinmat}\\
&&C_q(x)=\sum_{n=0}^{\infty}\,(-1)^n\,\frac{q^{n(n-1/2)}}{(q,q^2;q^2)_n}\,
x^{2n} \label{qcosmat}\, ,
\end{eqnarray}
where the notation $(a,b;q)_n$ used before means: $(a;q)_n(b;q)_n$
and we have used the identities
\begin{eqnarray}
\frac{(1-q)^{2n+1}}{(1-q)(q^2,q^3;q^2)_n}\,q^{n(n+1/2)}&=&\frac{1}{[2n+1]_q!} \,, \\
\frac{(1-q)^{2n}}{(q,q^2;q^2)_n}\,q^{n(n-1/2)}&=&\frac{1}{[2n]_q!}
\, .
\end{eqnarray}
Instead of the familiar identity $\sin^2x+\cos^2x=1$, one has
\cite{bustoz}
\begin{equation}
S_q(\q1 x)\, S_q(x)+C_q(\q1 x)\, C_q(x)=1  \, .
\end{equation}
Furthermore,
\begin{eqnarray}
&&\dj S_q (a\, x)=a\,C_q(a\, x) \, ,\\
&&\dj C_q (a\, x)=-a\,S_q(a\, x) \, ,
\end{eqnarray}
therefore, one can also easily see that $S_q(a\, x)$ and $C_q(a\,
x)$ are linearly independent solutions of the $q$-differential
equation
\begin{equation}
\dj^2 u(x)+a^2\, u(x)=0 \, . \label{wave_plane}
\end{equation}

The above introduced basic-trigonometric functions can be
expressed in terms of the basic-hypergeometric functions and of
the so-called Hahn-Exton $q$-Bessel function \cite{gasper,ismail}.
Furthermore, it is very relevant to observe that in
Ref.\cite{bustoz} it has been shown that $S_q(x)$ and $C_q(x)$
exhibit an analogue $q$-orthogonality relation and can be
considered Fourier expansions in these functions. Finally, it
should also be noticed that $q$-deformed polynomials, such as
$q$-deformed Hermite and $q$-deformed Laguerre polynomials have
been defined in literature in the framework of the
basic-hypergeometric functions \cite{rogers,vander,koe}.

\section{Operator properties in basic square-integrable space}
On the basis of the properties of the previous Section, we are
ready to introduce the main ingredients to develop a consistent
deformed quantum mechanics in the framework of $q$-calculus and
basic-hypergeometric functions.

Let $L^2_q$ be the basic square-integrable space of all complex
functions defined in $(-\infty,\infty)$ such that
\begin{equation}
\|\psi_q\|=\left( \int_{-\infty}^\infty \vert
\psi_q(x)\vert^2\,d_qx \right)^{1/2} <\infty\, .
\end{equation}
The space $L^2_q$ defined above, originally introduced in the
literature of the basic hypergeometric functions
\cite{exton,koe,annaby2003}, is a linear space. If $\psi_q$ and
$\varphi_q$ are two basic square-integrable functions, any linear
combinations $\alpha\psi_q+\beta\varphi_q$, where $\alpha$ and
$\beta$ are arbitrarily chosen complex numbers, are also basic
square-integrable functions. Moreover, $L^2_q$ is a separable
Hilbert space with the inner (scalar) product
\begin{equation}
\langle \varphi,\psi\rangle_q=\int \varphi^\star_q(x)\,\psi_q(x)\,
d_qx \, . \label{qscalar}
\end{equation}
A simple example of orthonormal basis of $L^2_q$ is (for $0<q<1$, in
the symmetric case) \cite{annaby2003}
\begin{equation}
\varphi_n(x)= \left\{
\begin{array}{rl}
\displaystyle
\frac{1}{\sqrt{x\,(\q1-q)}}\,, & \mbox{if } x=q^{2n+1} \, , \\
\\
0 \, , & \mbox{otherwise}\, ,
\end{array}
\right.
\end{equation}
with $n=\dots,-2,-1,0,1,2,\dots$. For completeness, let us mention
that there exist representations of the $q$-oscillator algebra
(\ref{algebra}) where the continuous $q$-Hermite polynomials play
the role of vector basis \cite{rogers,koe,floreanini}. Such
orthogonal polynomials can be expressed in terms of the
basic-hypergeometric functions and reduce, in the limit
$q\rightarrow 1$, to the standard Hermite polynomials.

The above scalar product (\ref{qscalar}) is linear with respect to
$\psi_q$, the norm of a function $\psi_q$ is a real, non-negative
number: $\langle \psi,\psi\rangle_q\ge 0$ and
\begin{equation}
\langle \psi,\varphi\rangle_q=\langle \varphi,\psi\rangle_q^\star
\, .
\end{equation}
Analogously to the undeformed case, it is easy to see that from the
above properties of the basic-scalar product follows the $q$-Schwarz
inequality
\begin{equation}
\vert \langle \varphi,\psi\rangle_q \vert^2 \le \langle
\varphi,\varphi\rangle_q\, \langle \psi,\psi\rangle_q \, .
\end{equation}

Consistently with the above definitions, the basic-adjoint of an
operator ${\hat A}_q$ is defined by means of the relation
\begin{equation}
\langle \psi,{\hat A}_q^\dag\varphi\rangle_q=\langle \varphi,{\hat
A}_q \psi\rangle_q^\star \, ,
\end{equation}
and, by definition, a linear operator is basic-Hermitian if it is
its own basic-adjoint. More explicitly, an operator ${\hat A}_q$ is
basic-Hermitian if for any two states $\varphi_q$ and $\psi_q$ we
have
\begin{equation}
\langle \varphi,{\hat A}_q \psi\rangle_q=\langle {\hat A}_q\varphi,
\psi\rangle_q \, .\label{hermi}
\end{equation}

On the basis of the results of the previous Section, it appears
natural to associate the operators corresponding to the position
coordinate and the momentum of a particle as follows
\begin{equation}
{\hat x}_q=x \,, \ \ \ {\hat p}_q=-i\hbar\dj . \label{xp_operator}
\end{equation}
At this point, it becomes a crucial question if the introduced
deformed momentum operator is basic-Hermitian. Let us therefore
consider the following product
\begin{equation}
\langle \varphi,\dj \psi\rangle_q= \int \varphi_q^\star(x)
\,\dj\psi_q(x)\, .
\end{equation}
By considering the integration by parts of Eq.(\ref{leibq}) and
imposing that the functions $\varphi_q(x)$ and $\psi_q(x)$ go to
zero in the limit of $x\rightarrow \pm\infty$, we have
\begin{eqnarray}
\langle \varphi,\dj \psi\rangle_q&=& -\int \dj\varphi_q^\star(q
x) \,\psi_q(q x)\, d_qx \nonumber \\
&=&  -\int {\cal D}_{qx} \varphi_q^\star(q x)\,\psi_q(q x)\, d_q(qx) \nonumber \\
&=&  -\int \dj \varphi_q^\star(y) \,\psi_q(y)\,  d_qy \, \, ,
\end{eqnarray}
where in the second equivalence we have used the property
(\ref{chain}) and in the last equivalence we have changed the
integration variable $y=qx$. Therefore, we have that the momentum
operator, introduced in Eq.(\ref{xp_operator}), is a basic-Hermitian
operator. Equivalently,
\begin{equation}
\langle \varphi,{\hat p}_q \psi\rangle_q=\langle {\hat p}_q\varphi,
\psi\rangle_q \, .\label{phermit}
\end{equation}
Let us remark that this important property is a peculiarity of the
$q\leftrightarrow\q1$ symmetric basic-framework defined in Eq.s
(\ref{algebra}), (\ref{bn}) and (\ref{jd}). Conversely, it is known
that, in the non-symmetric framework, the momentum operator defined
in term of the JD, like in Eq.(\ref{xp_operator}), is not Hermitian
\cite{epjc06,wess,annaby,lava08}.

\section{Observable and basic-deformed Schr\"odinger equation}
On the basis of the above properties, we have now the recipe to
generalize the de\-fi\-nition of an observable and to introduce a
basic-deformed Schr\"odinger equation in the framework of
$q$-deformed theory. Consistently with the standard quantum
mechanics, we have two fundamental postulates: 1) with the
dynamical variable $A(x,p)$ is associate the linear operator
${\hat A}_q(x,-i\hbar\dj)$; 2) the mean value of this dynamical
variable, when the system is in the (normalized) state $\psi_q$,
is
\begin{equation}
\langle {\hat A}\rangle_q=\int \psi^\star_q\, {\hat A}_q\,\psi_q
\,\,d_qx\, . \label{meanv}
\end{equation}

Observable are real quantities, hence the expectation value
(\ref{meanv}) must be real for any state $\psi_q$:
\begin{equation}
\int \psi^\star_q\,{\hat A}_q\,\psi_q \, d_qx=\int ({\hat
A}^\dag_q\,\psi^\star_q) \,\psi_q\, d_qx \, ,
\end{equation}
therefore, on the basis of Eq.(\ref{hermi}), observable must be
represented by basic-Hermitian operators.

If we require that there is a state $\psi_q$ for which the result
of measuring the observable $A$ is unique, in other words that the
fluctuations
\begin{equation}
(\Delta A_q)^2=\int \psi^\star_q\, ({\hat A}_q-\langle {\hat
A}\rangle_q)^2 \, \psi_q \, d_qx \, ,
\end{equation}
must vanish, we obtain the following basic-eigenvalue equation of a
basic-Hermitian operator $A_q$ with eigenvalue $a$
\begin{equation}
{\hat A}_q \,\psi_q= a \,\psi_q  \, .\label{eigen}
\end{equation}
As a consequence, the eigenvalues of a basic-Hermitian operator are
real because $\langle {\hat A}\rangle_q$ is real for any state; in
particular for an eigenstate with the eigenvalue $a$ for which
$\langle {\hat A}\rangle_q=a$.

Furthermore, as in the undeformed case, two eigenfunctions
$\psi_{q,1}$ and $\psi_{q,2}$ of the basic-Hermitian operator ${\hat
A}_q$, corresponding to different eigenvalues $a_1$ and $a_2$, are
orthogonal. We can always normalize the eigenfunction, therefore we
can chose all the eigenvalues of a basic-Hermitian operator
orthonormal, i.e.
\begin{equation}
\int \psi^\star_{q,n}\,\psi_{q,m} \, d_qx=\delta_{n,m} \, .
\end{equation}
Consequently, two eigenfunctions $\psi_{q,1}$ and $\psi_{q,2}$
belonging to different eigenvalues are linearly independent.

It easy to see that, adapting step by step the undeformed case to
the introduced $q$-deformed framework, the totality of the
linearly independent eigenfunctions $\{\psi_{q,n}\}$ of
basic-Hermitian operator ${\hat A}_q$ form a complete
(orthonormal) set in the space of the basic square-integrable wave
functions. In other words, if $\psi_q$ is any state of a system,
then it can be expanded in terms of the eigenfunctions (with a
discrete spectrum) of the corresponding basic-Hermitian operator
${\hat A}_q$ associate to the observable:
\begin{equation}
\psi_q=\sum_n c_{q,n}\, \psi_{q,n} \, ,
\end{equation}
where
\begin{equation}
c_{q,n}=\int \psi^\star_{q,n}\,\psi_q\, d_qx\, .
\end{equation}
The above expansion allows us, as usual, to write the expectation
value of ${\hat A}_q$ in the normalized state $\psi_q$ as
\begin{equation}
\langle {\hat A}\rangle_q=\int \psi^\star_q\, {\hat A}_q\,\psi_q
\,\,d_qx= \sum_n \vert c_{q,n}\vert^2_q \,\, a_n \, ,
\end{equation}
where $\{a_n\}$ are the set of eigenvalues (assumed, for
simplicity, discrete and non-degenera\-te) and the normalization
condition of the wave function can be written in the form
\begin{equation}
\sum_n \vert c_{q,n}\vert^2_q=1 \, .
\end{equation}

Following the prescriptions at the beginning of this Section, we are
able to introduce the basic-Hamiltonian operator as
\begin{equation}
{\hat H}_q=-\frac{\hbar^2}{2m}\, \dj^2+V_q(x) \, . \label{qhamil}
\end{equation}
The above Hamiltonian is basic-Hermitian and the time development of
a system will be given by the following basic-deformed Schr\"odinger
equation
\begin{equation}
i\hbar \frac{\partial \psi_q(x,t)}{\partial t}={\hat H}_q \,
\psi_q(x,t) \, .\label{qschro}
\end{equation}
The above equation implies a consistent conservation in time of
the probability density. In fact, by taking the complex
conjugation of Eq.(\ref{qschro}), summing and integrating term by
term the two equations, we get
\begin{equation}
i\hbar \frac{\partial}{\partial t}\int \psi^\star_q \, \psi_q \,
d_qx=\int \left[\psi^\star_q ({\hat H}_q\psi_q)-({\hat H}_q
\psi^\star_q)\psi_q \right ] \, d_qx =0\, ,
\end{equation}
where the last equivalence follows from the fact that the
Hamiltonian operator is basic-Hermitian.

The Eq.(\ref{qschro}) admits factorized solution
$\psi_q(x,t)=\phi(t)\,\varphi_q(x)$, where $\phi(t)$ satisfies to
the equation
\begin{equation}
i \hbar\frac{d \phi(t)}{d t}=E\phi(t) \, , \label{phit}
\end{equation}
with the standard (undeformed) solution
\begin{equation}
\phi(t)=\exp\left(-\frac{i}{\hbar}\,E\,t\right) \, ,
\end{equation}
while $\varphi_q(x)$ is the solution of time-independent
basic-Schr\"odinger equation
\begin{equation}
{\hat H}_q\, \varphi_q(x)=E\varphi_q(x) \, . \label{stationary}
\end{equation}

In one dimensional case, for a free particle ($V_q=0$) described by
the wave function $\varphi_q^f(x)$, Eq.(\ref{stationary}) becomes
\begin{equation}
\dj^2 \varphi_q^f(x)+k^2 \varphi_q^f(x)=0 \, ,
\end{equation}
where $k=\sqrt{2mE/\hbar^2}$. The previous equation is equivalent
to Eq.(\ref{wave_plane}) of Section II and, therefore, the
solution can be written in term of the basic-exponential
\begin{equation}
\varphi_q^f(x)=N\, {\rm E}_q(ikx)\, , \label{plane}
\end{equation}
or, equivalently, it can be expressed in term of a linear
combination of the basic-trigono\-me\-tric functions $S_q(kx)$ and
$C_q(kx)$, introduced in Eq.s (\ref{qsin}) and (\ref{qcos}). The
above equation generalizes the plane wave functions in our
framework of the $q$-calculus and basic hypergeometric functions.
In this context, as already remembered in the Introduction, it is
remarkable to note that the quantum dynamics of free Hamiltonian
and the properties of plane waves, eigenfunctions of the
$q$-deformed momentum operator, were also studied in Ref.s
\cite{wess2,kozyrev} in a different approach.

Moreover, by using the time-independent basic-Schr\"odinger
equation (\ref{stationary}), it is possible to study more complex
examples, such as the presence of the harmonic oscillator
potential. In this case, the eigenfunctions will be related to the
$q\leftrightarrow\q1$ symmetric $q$-Hermite polynomials
\cite{rogers,koe}. This important item will be extensively studied
in future investigations.

\section{Conclusion}
On the basis of the $q\leftrightarrow\q1$ symmetric calculus,
originally introduced in the framework of the basic-hypergeometric
functions, we have introduced a generalized linear Schr\"odinger
equation which involves a $q$-deformed Hamiltonian that is a
basic-Hermitian operator in the space of basic square-integral
wave functions. Although a complete description of the introduced
quantum dynamical equations lies out the scope of this paper, we
think that the results derived here appear to provide a starting
point to obtain a deeper insight into a full consistent
basic-deformed quantum mechanics in the framework of the
$q$-calculus and basic hypergeometric functions.

\end{document}